\documentclass[
prc,%
10pt,%
final,%
notitlepage,%
oneside,%
twocolumn,%
nobibnotes,%
nofootinbib,% In the current version of REVTeX, when this option is on,
superscriptaddress,%
floatfix,%
floatfix,%
showkeys,%
showpacs]%
{revtex4}
\usepackage{color}%\textcolor{red}{}
\usepackage{amsfonts}
\usepackage{amsbsy}
\usepackage{mathrsfs}
\usepackage{graphicx}
\def\lsim{\mathrel{\rlap{
\lower4pt\hbox{\hskip-3pt$\sim$}}
    \raise1pt\hbox{$<$}}}     %less than approx. symbol
\def\gsim{\mathrel{\rlap{
\lower4pt\hbox{\hskip-3pt$\sim$}}
    \raise1pt\hbox{$>$}}}     %greater than or approx. symbol

\begin{document}
\title{	
Global polarization in heavy-ion collisions based on axial vortical effect
} 
\author{Yu. B. Ivanov}\thanks{e-mail: yivanov@theor.jinr.ru}
\affiliation{Bogoliubov Laboratory for Theoretical Physics, 
Joint Institute for Nuclear Research, Dubna 141980, Russia}
\affiliation{National Research Nuclear University "MEPhI", %Kashirskoe sh. 31, 
Moscow 115409, Russia}
\affiliation{National Research Centre "Kurchatov Institute",  Moscow 123182, Russia} 
\begin{abstract}
Global polarization of $\Lambda$ 
and $\bar{\Lambda}$ is calculated based on the axial vortical effect (AVE). 
Simulations are performed within the model of the three-fluid dynamics. 
Equations of state with the   deconfinement transition result in a 
good agreement with STAR data for both $\Lambda$ and $\bar{\Lambda}$
polarization, in particular, with the  $\Lambda$-$\bar{\Lambda}$ splitting. 
Suppression of the gravitational-anomaly contribution required for the data
reproduction is in agreement with predictions of the QCD lattice simulations.
Predictions for the global polarization in forthcoming experiments at lower
collision energies are made. These forthcoming data will provide a 
critical test for the AVE and thermodynamic mechanisms of the polarization. 
\end{abstract}
\maketitle
%\today

\section{Introduction}
\label{Introduction}

Experimental discovery of global and local polarization of hadrons 
in STAR experiments \cite{STAR:2017ckg,Adam:2018ivw,Adam:2019srw} 
gave us evidence of existence of a new class of 
collective phenomena in heavy-ion collisions \cite{Becattini:2020ngo}.  
The thermodynamic approach based on hadronic degrees 
of freedom \cite{Becattini:2013fla,Becattini:2016gvu,Fang:2016vpj} 
well describes the global polarization of hyperons \cite{STAR:2017ckg,Adam:2018ivw}
as it was demonstrated by its implementations within  
various hydrodynamical 
\cite{Karpenko:2016jyx,Xie:2017upb,Ivanov:2019ern,Ivanov:2020wak}
and transport  
\cite{Li:2017slc,Wei:2018zfb,Shi:2017wpk,Kolomeitsev:2018svb,Vitiuk:2019rfv} 
models of heavy-ion collisions. 
However, this approach encounters problems.

The thermodynamic approach predicts 
wrong sign of the local longitudinal popularization as compared with that 
measured in the STAR experiment \cite{Adam:2019srw}. This discrepancy is rather robust, 
it comes out in both hydrodynamic \cite{Becattini:2017gcx,Xie:2016fjj}  
and transport \cite{Wei:2018zfb,Xia:2018tes,Wu:2019eyi} calculations. 
The thermodynamic approach fails to explain 
preliminary results on alignment of  of $\phi$ and $K^*$ mesons \cite{Singha:2020qns}
at energies of the Relativistic Heavy Ion Collider (RHIC).

The above problems indicate that the mechanism of particle polarization 
in heavy-ion collisions is not that clear so far. Therefore, alternative 
approaches should be considered. 
An alternative approach based on the axial vortical effect (AVE)
\cite{Vilenkin:1980zv,Son:2004tq,Sorin:2016smp}  
assumes equilibrium but not for spin degrees of freedom. 
The first applications of this approach \cite{Baznat:2017jfj,Sun:2017xhx} within the 
Quark-Gluon-String Model (QGSM) \cite{Toneev:1983cb,Toneev:1990vj,Amelin:1991cr} 
and a multiphase transport model \cite{Lin:2004en}
demonstrated its ability to describe the data on the global polarization
and to naturally explain  the  $\Lambda$-$\bar{\Lambda}$ splitting \cite{Baznat:2017jfj}. 
The AVE approach also gives qualitatively correct local longitudinal polarization
\cite{Liu:2019krs}.

In this paper, I report calculation of the global polarization of $\Lambda$ 
and $\bar{\Lambda}$ based on the AVE approach. 
Simulations are performed within 3FD model \cite{3FD}. The 3FD model is based on 
a minimal way to implement the early-stage nonequilibrium of the produced 
strongly-interacting matter at high collision energies.
This early nonequilibrium stage 
is modeled by means of two counterstreaming baryon-rich fluids (p and t fluids). 
Newly produced particles, dominantly
populating the midrapidity region, are associated with a fireball (f) fluid.
These fluids are governed by conventional hydrodynamic equations 
coupled by friction terms in the right-hand sides of the Euler equations. 

Calculations were carried out  
with two versions of equation of state (EoS) with the  deconfinement
 transition \cite{Toneev06}, i.e. a first-order phase (1PT) transition  
and a crossover one. 
Results with hadronic EoS \cite{Mishustin:1991sp} are also presented. 
The physical input of the present 3FD calculations is described in
Ref.~\cite{Ivanov:2013wha}.

\section{Polarization based on the AVE} 
\label{AVE}

Presence of vorticity in a system 
   \begin{eqnarray}
\label{vorticity}
   \omega_{\mu\nu} = \frac{1}{2}
   (\partial_{\nu} u_{\mu} - \partial_{\mu} u_{\nu}),  
   \end{eqnarray}
where $u_{\mu}(x)$ is local 4-velocity of the medium, 
induces the axial current of chiral particles 
\begin{equation}
\label{J5}
J^\nu_{5}(x)=N_c %\int d^3 x \,
\left(\frac{\mu^2}{2 \pi^2}+\kappa\frac{T^2}{6}\right) 
{\epsilon^{\nu\alpha\beta\gamma}u_{\alpha}
\partial_{\beta}u_\gamma}
\end{equation}
where $\mu$ is the chemical potential of these particles, 
 $T$ is the temperature of the medium, and $\kappa$ is a parameter discussed below. 
Spins of these particles get aligned along the direction of the axial current. 
Thus, these particles become polarized. This is the essence of the axial vortical effect
\cite{Vilenkin:1980zv,Son:2004tq,Sorin:2016smp}.

While the first term $\sim\!\!\mu^2$ in the braces of Eq. (\ref{J5}) is topologically 
protected, i.e. it is related to topological invariant in the momentum space, the 
term $\sim\!\!\ T^2$ related to gravitational anomaly \cite{Landsteiner:2011iq} is not. 
Therefore, similarly to Ref. \cite{Baznat:2017jfj} a 
parameter $\kappa$ is introduced into Eq. (\ref{J5}) which scales this 
gravitational term. 
Lattice simulations of Ref. \cite{Braguta:2013loa} result in
zero $\kappa$ in the confined phase and 
predict suppression of the gravitational term by one order of magnitude, 
i.e. $\kappa\approx 0.1$, at very high temperatures $T>400$ MeV.

I am interested in $\Lambda$ and $\bar{\Lambda}$ hyperons. 
Their polarization is related to 
the  axial current of (anti)strange quark $J^\nu_{5s}$, 
which differs from (\ref{J5}) by replacement 
of the chemical potential $\mu$ by the chemical potential of (anti)strange quark 
$\mu_s = -\mu_{\bar{s}} = \mu_B/3 - \mu_S$, where $\mu_B$ is the baryon chemical potential
and  $\mu_S$ is the strange one.
%The polarization of strange and anti-strange quarks is related to $J^\nu_{5s}$.  
Following Refs. \cite{Sorin:2016smp,Baznat:2017jfj}, 
the global polarization of $\Lambda$ and $\bar{\Lambda}$ 
%is associated with the polarization 
%of strange and anti-strange quarks, respectively, and 
is related to $J^\nu_{5s}$ as 
\begin{eqnarray} 
\label{PL}
P_{\Lambda}  
%&=&  \left\langle \frac{m_{\Lambda}}{(N_{\Lambda}+N_{\bar{K}^*})p_y } 
%J^0_{5s} \right\rangle, 
&=&  \int d^3 x \, (J^0_{5s}/u_y) /(N_{\Lambda}+N_{\bar{K}^*}), 
\\
\label{PaL}
P_{\bar{\Lambda}} 
%&=&  \left\langle \frac{m_{\bar{\Lambda}}}{(N_{\bar{\Lambda}}+N_{K^*})p_y } 
%J^0_{5s} \right\rangle
&=&  \int d^3 x \, (J^0_{5s}/u_y) /(N_{\bar{\Lambda}}+N_{K^*}), 
\end{eqnarray}
where 
%$p_y$ is $\Lambda$ or $\bar{\Lambda}$ transverse momentum, 
$u_y$ is $y$ component of the 4-velocity, 
$N_{\Lambda}$ and $N_{\bar{\Lambda}}$ are numbers of produced  
$\Lambda$'s and $\bar{\Lambda}$'s, respectively, and
$N_{K^*}$ and $N_{\bar{K}^*}$ are numbers of produced  
$K^*$ and $\bar{K}^*$ mesons, respectively.  
%$m_{\Lambda}=m_{\bar{\Lambda}}$ the mass of $\Lambda$ and $\bar{\Lambda}$. 
Here $N_{\Lambda}$ and $N_{\bar{K}^*}$ count the number of strange quarks 
which carry the polarization, similarly for anti-strange quarks. 
This is because only strange particles with nonzero spin  carry the $s$-quark polarization. 
The $1/u_y$ factor results from boost to the rest frame of the fluid element. 
%where the polarization is measured.  
%The $1/p_y$ factor results from boost to the rest frame of the produced hyperon, 
%where the polarization is measured.  

Expressions (\ref{PL}) and (\ref{PaL}) are just estimates of the polarization 
rather than rigorously derived formulas. 
In the original papers 
\cite{Sorin:2016smp,Baznat:2017jfj} the boost was made to the rest frame 
of the produced hyperon, where the polarization is measured.  
A shortcoming of that recipe is that 
the result of averaging over momenta of produced hyperons diverges. 
Indeed, only the $1/p_y$ factor, resulted from that boost, 
depends on the momentum in Eqs. (\ref{PL}) and (\ref{PaL}), 
which results in divergence at low $p_y$. 
In addition, the boost to the fluid local rest frame is more natural because  
this approach deals only with properties 
of the medium rather than with separate particles.  
The $u_y$ component of the 4-velocity in Eq. (\ref{J5}) cancel 
the $1/u_y$ factor thus eliminating the divergence.

Thus, collecting all together, I arrive at the expression 
in terms quantities averaged over the medium ($\langle ...\rangle$)
\begin{equation}
\label{polarization} 
P_{\Lambda}  
%=  
%\frac{1}{(n_{\Lambda}+n_{\bar{K}^*})}
%N_c \langle \left(\frac{\mu_s^2}{2 \pi^2}+\kappa\frac{T^2}{6}\right) 
%{\epsilon^{02\beta\gamma}
%\partial_{\beta}u_\gamma}\rangle
= 
 \frac{N_c}{\langle n_{\Lambda}+n_{\bar{K}^*} \rangle}
 \left\langle \frac{\mu_s^2}{2 \pi^2}+\kappa\frac{T^2}{6} \right\rangle 
%\epsilon^{0yxz}
\langle \omega_{xz}\rangle
\end{equation}
where $n_{\Lambda}$ and $n_{\bar{K}^*}$ are  
densities of $\Lambda$'s and $\bar{K}^*$ mesons, respectively. 
Similar result holds for $P_{\bar{\Lambda}}$.  
Here I decoupled averaging of the vorticity and the prefactor.

\section{Global polarization}
\label{Global polarization}

The above approach is very suitable for the calculation of the global vorticity 
within the method suggested in Refs. \cite{Ivanov:2019ern,Ivanov:2020wak}. 
This method consists in calculation of average polarization in the central region 
of colliding nuclei, the right and left borders of which are chosen from 
the condition $|y|\lsim 0.5$. The rapidity $y$ is calculated based on hydrodynamical 
velocities. The experimental acceptance $|\eta|<1$, where $\eta$ is pseudorapidity, 
better comply with the condition $|y|\lsim 0.7$ in terms of the true $y$ rapidity
\cite{Ivanov:2020wak}. 
However, hydrodynamical rapidity does not well coincide with the true one 
at low collision energies. 
Taking also into account that the condition $|y|\lsim 0.5$ results in better 
reproduction of the data at low collision energies while only slightly differing 
from the $|y|\lsim 0.7$ results at high energies \cite{Ivanov:2020wak}, 
I took the condition $|y|\lsim 0.5$ for the theoretical acceptance. 
Details of the polarization calculation in the central region are described in 
Ref. \cite{Ivanov:2020wak}. The calculation of the AVE polarization
is very similar to the thermodynamical one in Ref. \cite{Ivanov:2020wak}. 
3FD simulations of Au+Au collisions were performed at fixed
impact parameter $b=$ 8 fm. This impact parameter was taken to roughly 
comply with the STAR centrality selection of 20-50\% \cite{STAR:2017ckg}. 
Glauber simulations of Ref. \cite{Abelev:2008ab} were used to relate 
the experimental centrality and the mean impact parameter. 

%\textcolor{red}{
In fact, the present evaluation of the global polarization is an estimate rather
than a calculation, as it is discussed in Ref. \cite{Ivanov:2020wak} in more detail. 
This is the result of approximations made:  
the already mentioned decoupling of averaging of the vorticity and the prefactor  
in Eq. (\ref{polarization}), the isochronous freeze-out instead of the local one 
conventionally used in the 3FD, the above-mentioned hydrodynamical-rapidity window 
instead of the true experimental acceptance. Some of these approximations 
were tested in Refs. \cite{Ivanov:2019ern,Ivanov:2020wak} within the thermodynamic approach. 
Since the estimate of the global polarization within the AVE approach is technically 
very similar to that within the thermodynamic approach,  
uncertainties are also very similar --- the overall uncertainty generally
does not exceed 20\%.
%}

%
%\begin{figure}[p]
\begin{figure}[bht]
%\vspace*{8mm}
\includegraphics[width=8.0cm]{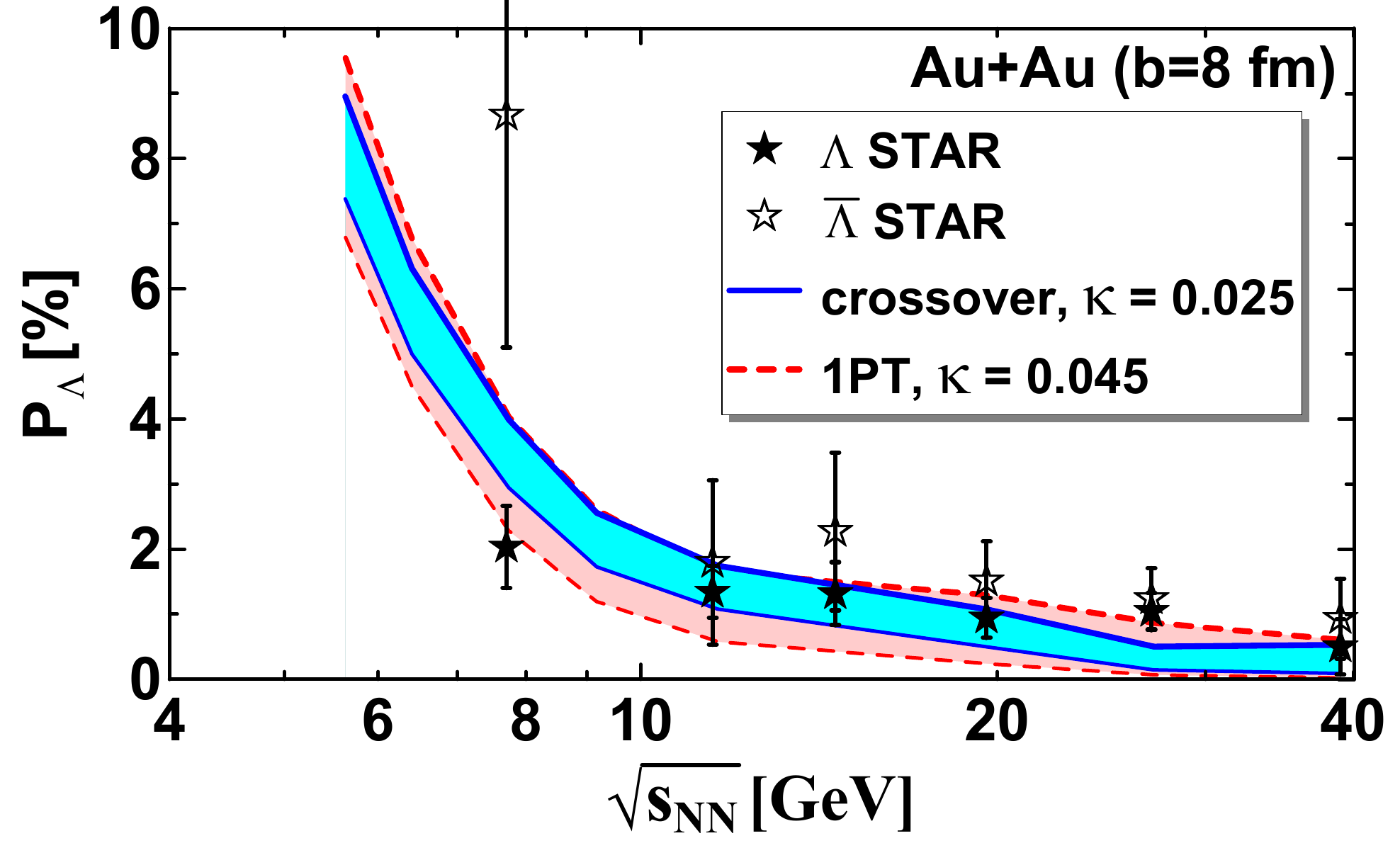}
 \caption{%(Color online)
 Global polarization of $\Lambda$ hyperons in Au+Au collisions at $b=$ 8 fm as 
function of collision energy $\sqrt{s_{NN}}$.  
Upper borders of the bands correspond to parameters $\kappa$ [see Eq. (\ref{J5})] 
displayed in the legend, while the lower borders -- to $\kappa=0$. 
STAR data on global $\Lambda$ and $\bar{\Lambda}$ 
polarization \cite{STAR:2017ckg} are also displayed. 
}
\label{fig1}
\end{figure}
%
%
%\begin{figure}[p]
\begin{figure}[bht]
%\vspace*{8mm}
\includegraphics[width=8.0cm]{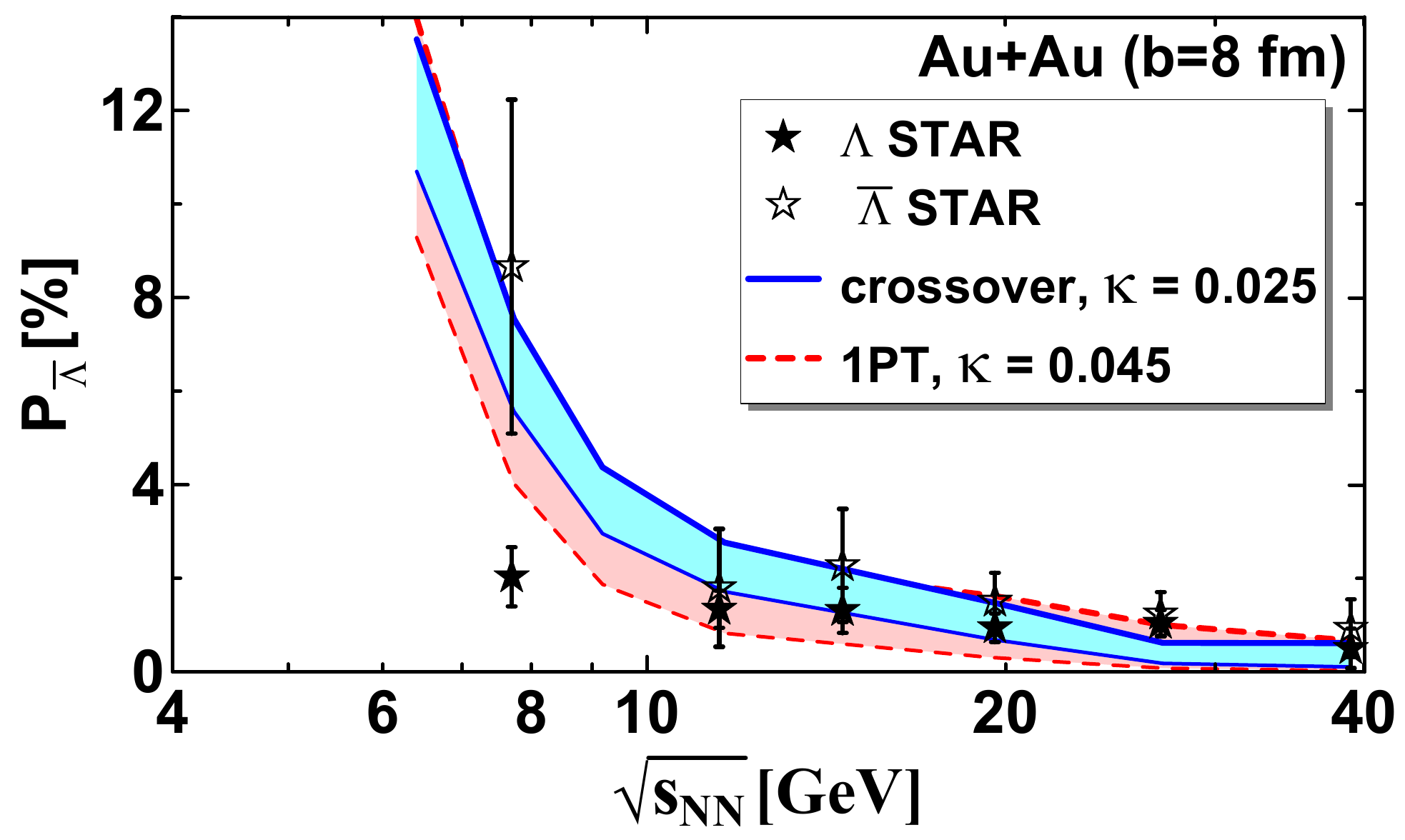}
 \caption{%(Color online)
The same as in Fig. \ref{fig1} but for the $\bar{\Lambda}$ hyperons. 
}
\label{fig2}
\end{figure}

The obtained results are presented in Figs. \ref{fig1} and \ref{fig2}.  
In view of above said about the gravitational term $\sim\!\!\ T^2$, I present 
results without this term, i.e.  $\kappa=0$, see Eq. (\ref{J5}), and with this 
term fitted to reproduce data on the global polarization at high collision energies. 
The results of these fitted values of $\kappa$ are displayed 
in Figs. \ref{fig1} and \ref{fig2}. Though being EoS dependent, the matched values of 
$\kappa$ are the same for $\Lambda$'s and $\bar{\Lambda}$'s, as it should be.
Moreover, at lowest considered collision energy, $\sqrt{s_{NN}}=$ 7.7 GeV,  
the data \cite{STAR:2017ckg} better agree with results without the gravitational term, 
i.e.  $\kappa=0$, while at higher energies - with those, where 
this term is suppressed by more than one order of magnitude. 
All this agrees with predictions of 
lattice simulations of Ref. \cite{Braguta:2013loa}. 
Suppression by more than one order of magnitude at higher energies 
also matches with the lattice results \cite{Braguta:2013loa}.

%
%\begin{figure}[p]
\begin{figure}[bht]
%\vspace*{8mm}
\includegraphics[width=8.0cm]{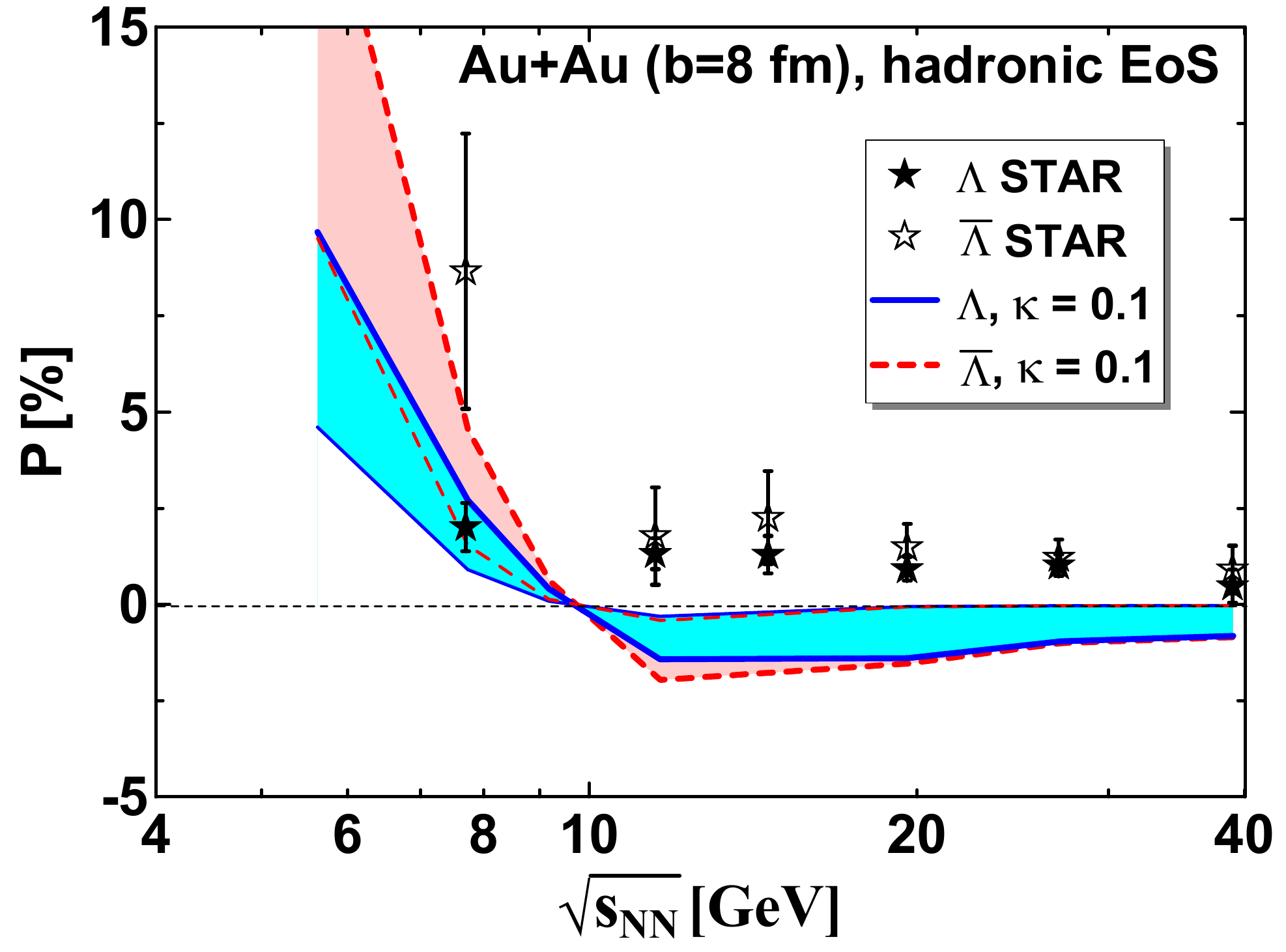}
 \caption{%(Color online)
 Global polarization of $\Lambda$ and $\bar{\Lambda}$ 
hyperons in Au+Au collisions at $b=$ 8 fm as 
function of collision energy $\sqrt{s_{NN}}$ in simulations 
with hadronic EoS of Ref. \cite{Mishustin:1991sp}.  
Bold borders of the bands correspond to parameters $\kappa$ [see Eq. (\ref{J5})] 
displayed in the legend, while the thin borders -- to $\kappa=0$. 
STAR data on global $\Lambda$ and $\bar{\Lambda}$ 
polarization \cite{STAR:2017ckg} are also displayed. 
}
\label{fig3}
\end{figure}

Not any EoS results in reasonable agreement with the STAR data on the global polarization. 
The purely hadronic EoS of Ref. \cite{Mishustin:1991sp} fails to reproduce the data, 
see  Fig. \ref{fig3}. The parameters $\kappa=$ 0.1 is chosen just to comply with the data 
at the lowest measured energy of 7.7 GeV.  
%\textcolor{red}{
Contrary to the data, the hadronic-EoS 
polarization becomes even negative at higher collision energies
because the vorticity $\langle \omega_{xz}\rangle$ in Eq. (\ref{polarization})  
becomes negative. The reason is that the vorticity in the central region does not 
correlate with the angular momentum \cite{Ivanov:2020wak} at the freeze-out stage,
while initially the angular momentum is the driving force for the vortex generation.
%}
This is similar to the 
polarization within the thermodynamic approach, which is discussed in detail in 
Ref.  \cite{Ivanov:2020wak}. 
The hadronic EoS also fails to reproduce a number of bulk observables at high energies, 
while the crossover and 1PT EoS's describe bulk observables equally good, as a rule. 
Thus, there is a correlation between reproduction of the global polarization 
and other bulk and flow observables.

%\vspace*{3mm} 
\section{Conclusions}
\label{Conclusions}

Calculation of the global polarization of $\Lambda$ 
and $\bar{\Lambda}$ are made based on the AVE approach. 
Simulations are performed within 3FD model \cite{3FD}. 
EoS's with the   deconfinement transition result in a 
good agreement with STAR data \cite{STAR:2017ckg}  
for both $\Lambda$ and $\bar{\Lambda}$
polarization, in particular, with the  $\Lambda$-$\bar{\Lambda}$ splitting. 
%\textcolor{red}{
This result is a consequence of the assumption made for the 
hadronization \cite{Sorin:2016smp}. 
It is assumed that the strange axial charge is conserved at the hadronic level after hadronization in spite of a very approximate chiral symmetry in the strange sector. 
In principle, an alternative assumption is possible which  
is based on the coalescence model for the hadronization: quarks coalesce into hadrons, keeping their polarization. Then the $\Lambda$ and $\bar{\Lambda}$  polarizations of would be very close to each other and would differ only due to nuclear-medium 
\cite{Vitiuk:2019rfv,Csernai:2018yok} 
and/or electromagnetic effects \cite{Guo:2019joy,Han:2017hdi}, similarly to that within the thermodynamic approach \cite{Becattini:2013fla,Becattini:2016gvu,Fang:2016vpj}.  
%}

Suppression of the gravitational-anomaly contribution required for the data
reproduction is in agreement with predictions of the QCD lattice 
simulations \cite{Braguta:2013loa}.
At the lowest considered collision energy, $\sqrt{s_{NN}}=$ 7.7 GeV,  
the data  better comply with results without the gravitational term, 
while at higher energies - with results, where 
this term is suppressed by more than one order of magnitude. 
At the same time, the hadronic EoS fails to reproduce the data on the global polarization.

The AVE global polarization rises with the collision-energy decrease faster than 
the thermodynamic polarization does \cite{Ivanov:2020wak}, as the 3FD simulations 
beyond the RHIC range indicate. Therefore, the forthcoming data from 
the Facility for Antiproton and Ion Research (FAIR)
and Nuclotron-based Ion Collider fAcility (NICA) will provide a critical test for 
the AVE and thermodynamic approaches.

\vspace*{2mm} 
{\bf Acknowledgments} \vspace*{1mm}

%\begin{acknowledgments} 
Fruitful discussions with O.V. Teryaev and D.N. Voskresensky are gratefully acknowledged.
This work was carried out using computing resources of the federal collective usage center ``Complex for simulation and data processing for mega-science facilities'' at NRC "Kurchatov Institute", http://ckp.nrcki.ru/.
This work was partially supported by the Russian Foundation for
Basic Research, Grants No. 18-02-40084 and No. 18-02-40085,  
%A.A.S. was partially supported 
and by  the Ministry of Education and Science of the Russian Federation within  the Academic Excellence Project of the NRNU MEPhI under contract No. 02.A03.21.0005.  
%\end{acknowledgments}

\end{document}